\documentclass[aps,prl,reprint,superscriptaddress]{revtex4-1}

\usepackage{harpoon}
\usepackage{float}
\usepackage{appendix}
\usepackage{mathptmx}
\usepackage{color}
\usepackage{upgreek}
\usepackage{graphicx}
\usepackage[colorlinks, citecolor=blue, linkcolor=blue, urlcolor=blue]{hyperref}
\usepackage{amssymb}
\usepackage{latexsym}
\usepackage{multirow}
\usepackage{amsmath}
\usepackage{siunitx}
\usepackage[normalem]{ulem}
\usepackage{multirow,makecell,diagbox}
\usepackage{booktabs}

\begin{document}

	\title{Visualizing Three-Dimensional Quantum Fluctuations of Ground-State C$_3$H$_4$ via Deep Sub-Angstrom Coulomb Explosion Imaging}

	\author{Mei Wang}
	\author{Yue Gao}
	\affiliation{State Key Laboratory of Heavy Ion Science and Technology, Institute of Modern Physics, Chinese Academy of Sciences, Lanzhou 730000, China} 
	\affiliation{School of Nuclear Science and Technology, University of Chinese Academy of Sciences, Beijing 100049, China}
	
	\author{Hang Yuan}
	\affiliation{State Key Laboratory of Heavy Ion Science and Technology, Institute of Modern Physics, Chinese Academy of Sciences, Lanzhou 730000, China} 
	
	\author{Yang Wang}
	\affiliation{School of Chemistry and Materials, Yangzhou University, Yangzhou, 225002, China}
	\affiliation{Jiangsu Provincial Key Laboratory of Green \& Functional Materials and Environmental Chemistry, Yangzhou University, Yangzhou 225002, Jiangsu, China}
	
	\author{Bo Yang}
	\author{Shaofei Gu}
	\affiliation{State Key Laboratory of Heavy Ion Science and Technology, Institute of Modern Physics, Chinese Academy of Sciences, Lanzhou 730000, China} 
	\affiliation{School of Nuclear Science and Technology, University of Chinese Academy of Sciences, Beijing 100049, China}
	
	\author{Shaofeng Zhang}
	\author{Xinwen Ma}
	\author{Shenyue Xu}
	\email[]{s.xu@impcas.ac.cn}
	\affiliation{State Key Laboratory of Heavy Ion Science and Technology, Institute of Modern Physics, Chinese Academy of Sciences, Lanzhou 730000, China} 
	\affiliation{School of Nuclear Science and Technology, University of Chinese Academy of Sciences, Beijing 100049, China}

	\date{\today}
	\begin{abstract}

Coulomb explosion imaging (CEI) is a powerful tool for determining static geometries of isolated gas-phase molecules. In this work, taking allene and propyne as prototype, we demonstrate sub-angstrom resolution achieved with highly charged ion-induced CEI. Using 41.9 keV/u Xe$^{20+}$ beam, minute variations of CC single, double and triple bond lengths are clearly visualized, indicating that an ultrahigh resolution of 0.1 $\mathring{\mathrm{A}}$ is achieved. This unprecedented resolution not only makes it possible to distinguish CC bonds with different bond orders, but also lays a solid foundation for direct observation of quantum fluctuations inherent to 3D structure of a ground-state molecule. Notably, the coupled collective motion of the two CH$_2$ moieties spaced by a central carbon within allene is unambiguously visualized in our measurements.

	\end{abstract}
	
	\maketitle

%\linenumbers

Atomic-scale spatial resolution, especially sub-angstrom precision down to 0.1 $\mathring{\mathrm{A}}$, is a pivotal yet technically challenging capability for physics, chemistry and material researches \cite{Nellist2004, Jiang2018, Blackburn2025}. In the design of functional materials, minute atomic displacements can remarkably alter the intrinsic physicochemical properties of materials \cite{Masar2024}. At an ultrahigh precision of 0.1 $\mathring{\mathrm{A}}$, subtle structural features in molecular geometry can be visualized. These include bond length variations among covalent bonds with distinct bond orders, as well as structural variations arising from intrinsic quantum fluctuations \cite{Richard2025}. A series of advanced imaging technologies have been developed to access atomic level structural resolution. For example, atomic force microscopy \cite{Binnig1986} and electron microscopes \cite{Williams1996, Seiler1983} provide atomic resolution to the surface of solid and thin film materials \cite{Gross2009, Telychko2022, Qiu2025}. Electron diffraction \cite{Hensley2012, Wolter2016, Yang2018, Wolf2019} and X-ray diffraction \cite{Kupper2014, Minitti2015, Stankus2019} were initially developed for crystals, and are now extended to gas-phase molecules \cite{Hensley2012, Wolter2016, Yang2018, Wolf2019, Kupper2014, Ma2022}. Nevertheless, only the orientation-averaged structural information can be obtained by these diffraction methods, and the sensitivity to light elements like hydrogen is inherently restricted by the small scattering cross-sections \cite{Odate2023}.

Coulomb explosion imaging (CEI) \cite{Vager1989} is a mass spectrometry-based technique ideally suited for imaging the 3D structures of gas-phase molecules. It exhibits high and equal sensitivity to both hydrogen and the other heavier elements. The key prerequisites to achieve high precision CEI is to instantly strip multiple electrons from the molecule, thereby charging up all consisting atoms prior to structural rearrangement or relaxation \cite{Zhou2020, Yuan2024}. Recent works based on X-ray pulses or intense laser pulses have successfully achieved clear structural imaging of complex molecules \cite{Richard2025, Boll2022, Lam2024, Green2025, Cheng2026}. Nevertheless, inherent to the duration of these laser pulses as well as the Auger decay mechanisms following X-ray absorption, significant deviation appears between the measured momenta of the fragments and calculated momenta taking neutral structure of the molecule as initial conditions \cite{Boll2022, Lam2024}. This impedes the direct extraction of high resolution structural information. As a compromise, theoretical methods have been developed to extract molecular geometry, either by detailed modeling of fragmentation mechanisms \cite{Cheng2026, Richard2021, Ashrafi2024, Viveiros2026} or by employing machine learning to establish empirical correlations between measured momenta and real-space geometry \cite{Venkatachalam2025, Guo2025, Li2026, Andre2025}.
	
Swift highly charged ion (HCI) can trigger fast and violent ionization upon collision. In such a process, a multitude of electrons can be removed on an ultrashort timescale of approximately 100 as, primarily from the outer-valence shell of the target \cite{Yuan2024}. This feature makes HCI an ideal tool for high precision CEI of static structures. Recent work taking planar heterocyclic C$_4$H$_4$N$_2$ molecules as a prototype, has demonstrated that high precision momentum imaging can be achieved equally for both hydrogen and heavier atoms via HCI \cite{Yuan2024}. Compared with aromatic molecules possessing rigid ring frameworks, acyclic molecules with 3D structures are more flexible, leading to a broader fluctuation range of ground-state structures. The precision of HCI-induced CEI in stereo molecules remains to be verified. In the present study, we take two isomers of C$_3$H$_4$, allene (CH$_2$CCH$_2$) and propyne (CH$_3$CCH) as a prototype, and demonstrate that an unprecedented deep sub-angstrom resolution can be achieved employing HCIs. Crucially, this ultrahigh resolution enables direct visualization of quantum fluctuations within the full 3D ground-state molecular geometries. Such an experimental observation has only been reported very recently, and holds valid exclusively under specific geometric constraints \cite{Richard2025}.
	
The experiment was performed employing the reaction microscope \cite{Ullrich2003} (also called cold target recoil ion momentum spectroscopy \cite{Dorner2000}) mounted on the 320 kV platform of the Heavy Ion Research Facility in Lanzhou (HIRFL). More experimental details are given in the Supplemental Material Sec. I \cite{supp} and Refs. \cite{Yuan2024, Gao2025}. Briefly, we used 41.9 keV/u Xe$^{20+}$ beam to collide with C$_3$H$_4$ jet to trigger Coulomb explosion. The target jet was produced by supersonic expansion of pure C$_3$H$_4$ gas at room temperature with stationary pressure of 0.5 bar through a nozzle with a diameter of 60 $\mu$m. Under such conditions, the molecules remain predominantly in the vibrational ground-state (see Supplemental Material Sec. II \cite{supp} and Refs. \cite{Miller1988, Siddiki2022} for details). After collision, several ionic fragments are recorded in coincidence with one emitted electron, and the momentum vectors of these fragments are reconstructed. We choose the 2H$^+$ + 3C$^+$ coincidence events to extract structural information.

	%%%%%%%%% fig 1 %%%%%%%%%%%%%%%%%%%%%%%%%%%%%%%%%%%%%
	\begin{figure}[t]
		\centering
		\includegraphics[width=1 \columnwidth]{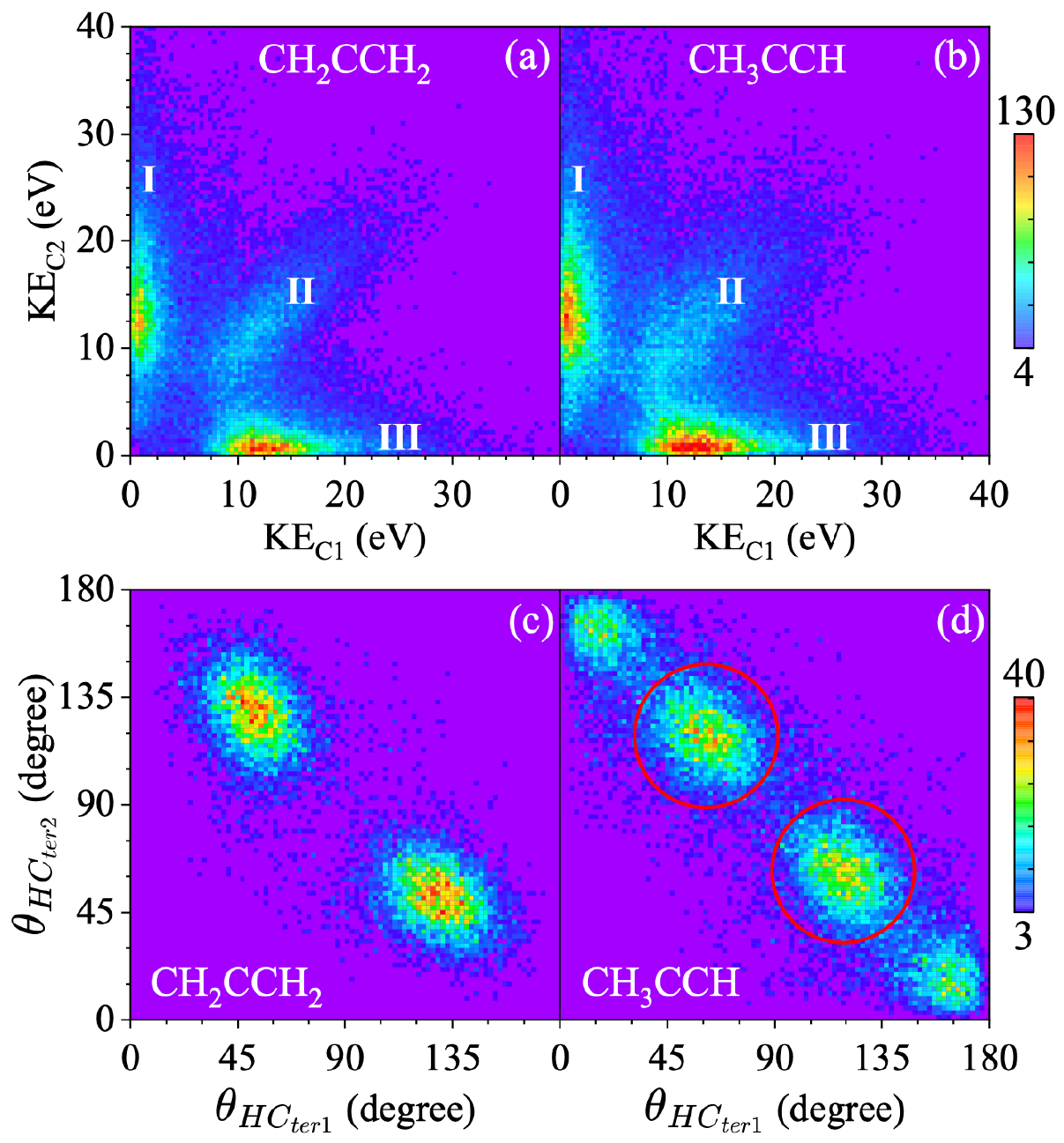}
		\caption{
Identifying terminal C atoms and H atoms attached to them from KE and mutual angle correlations. (a) and (b) KE correlation spectra between two C$^+$. Three isolated islands appear in these plots as marked by I, II and III. (c) and (d) angular correlation spectra between $\theta_{HC_{ter1}}$ and $\theta_{HC_{ter2}}$. Here $\theta_{HC_{ter1}}$ and $\theta_{HC_{ter2}}$ are defined as mutual angles between momentum vectors of H$^+$ and terminal C$^+$. 
			\label{fig1}}
	\end{figure}
	%%%%%%%%%%%%%%%%%%%%%%%%%%%%%%%%%%%%%%%%%%%%%%%%%

Figures 1(a) and (b) display the kinetic energy (KE) correlation between two of the three C$^+$ ions for allene and propyne, respectively. Three isolated structures marked as I, II and III appear in both plots. In region I and III, the KE distribution of one C$^+$ mainly ranges from 8 eV to 22 eV, while the other C$^+$ peaks near zero, indicating that one C$^+$ locates at terminal position of the skeleton while the other locates at the center. In region II, the two C$^+$ ions exhibit comparable KEs ranging from 8 eV to 22 eV, both assigned to terminal carbons. The two terminal C$^+$ ions for CH$_2$CCH$_2$ are identical to each other since both of them locates at the CH$_2$ group. While for CH$_3$CCH the two terminal C$^+$ are not identical. The affiliated group for each H and C can be identified in the angular correlation spectra shown in Figs. 1(c) and (d). Here the $\theta_{HC_{ter1}}$ and $\theta_{HC_{ter2}}$ are defined as the angles between momentum vectors of each H$^+$ and the two terminal C$^+$ ions. For CH$_2$CCH$_2$, two islands with mutual angles of \SI{50}{\degree}  and \SI{130}{\degree}  appear, corresponding to the proton either from the same CH$_2$ group with the detected C$^+$ or from the opposite CH$_2$ group, respectively. For CH$_3$CCH, the two islands with mutual angles of \SI{60}{\degree}  and \SI{120}{\degree}  (red circles) correspond to H$^+$ from CH$_3$ group and C$^+$ from either CH$_3$ group or CH group on the opposite side, respectively. The islands with mutual angles of \SI{15}{\degree}  and \SI{165}{\degree}  are distinct characteristic of H$^+$ from CH group.

%%%%%%%%% fig 2 %%%%%%%%%%%%%%%%%%%%%%%%%%%%%%%%%%%%%
	\begin{figure}[b]
		\centering
		\includegraphics[width=1 \linewidth]{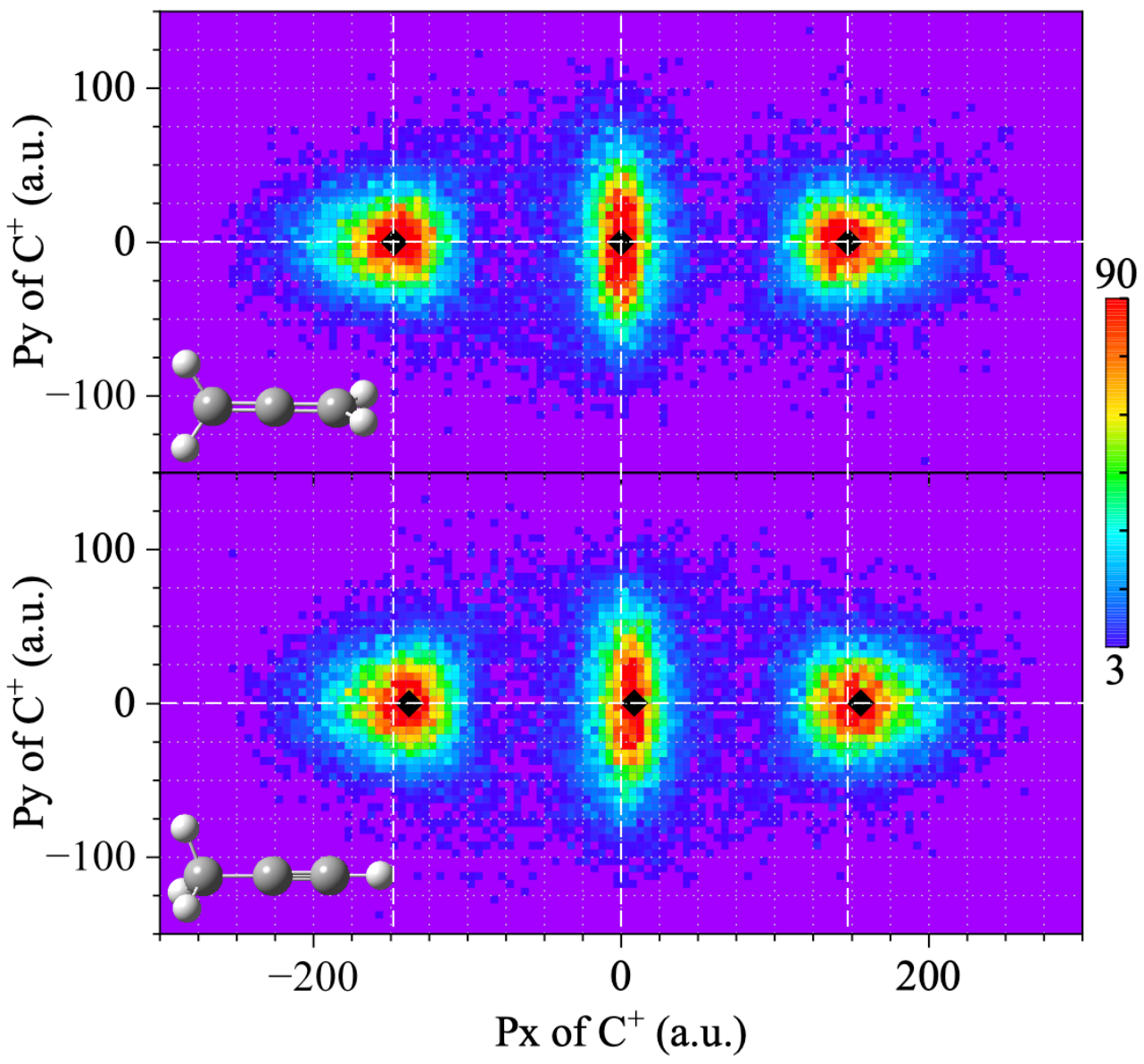}
		\caption{
Momentum imaging of the carbon skeleton shown in Newton diagrams. (a) for CH$_2$CCH$_2$ and (b) for CH$_3$CCH. Experimental data are compared with PC calculations (black diamonds). 
			\label{fig2}}
	\end{figure}
	%%%%%%%%%%%%%%%%%%%%%%%%%%%%%%%%%%%%%%%%%%%%%%%%%%%%%

We now define the molecular frame and image the carbon skeleton.
\textcolor{black}{The x-axis in the molecular frame is defined as the momentum-difference vector of the two terminal C$^+$ ions.}
For CH$_2$CCH$_2$, we randomly select one of the two detected H$^+$ to define the xy plane, and rotate this H$^+$ randomly into the second or third quadrant. For CH$_3$CCH, one detected H$^+$ from CH$_3$ group is selected to define the xy plane, and its momentum vector is rotated to the second quadrant of xy plane (see Supplemental Material Sec. III for details \cite{supp}). Figure 2 presents the measured momentum distributions, the so-called Newton diagrams, for C$^+$ ions in the defined molecular frame. Black diamonds represent point-charge (PC) approximation calculations taking the geometries of neutral molecule as initial conditions (see Supplemental Material Sec. IV \cite{supp} and Refs. \cite{NISTWebbook, Gaussian16, Mebel1998, Psciuk2010} for details). Both isomers show overall similar profiles where the central C$^+$ is well resolved from the two terminal C$^+$, and fit well with PC calculations. Notably, close comparison reveals subtle distinctions in momentum distributions between the two isomers. For CH$_2$CCH$_2$, the momentum of the central carbon peaks nearly zero, reflecting the symmetric CCC backbone and identical CH$_2$ configurations on both sides. In contrast, for CH$_3$CCH the momenta of the central carbon and two terminal carbons shift slightly toward +x direction (toward CH group), yielding an overall asymmetric profile. Such asymmetry directly reflects the unequal inter-nuclear distances of the CC single and double bonds and the asymmetric H configurations. Since the lengths of CC single, double and triple bonds in the two isomers varies merely 0.1 $\mathring{\mathrm{A}}$, \textcolor{black}{the distinct capture of such subtle difference} proves direct evidence that deep sub-angstrom resolution can be achieved with HCI beam. In addition, a noticeably broader y-distribution is observed for the central C$^+$. As we will demonstrate later, such features contain essential information regarding the ground-state quantum fluctuations.

	%%%%%%%%% fig 3 %%%%%%%%%%%%%%%%%%%%%%%%%%%%%%%%%%%%%
	\begin{figure}[b]
		\centering
		\includegraphics[width=1 \columnwidth]{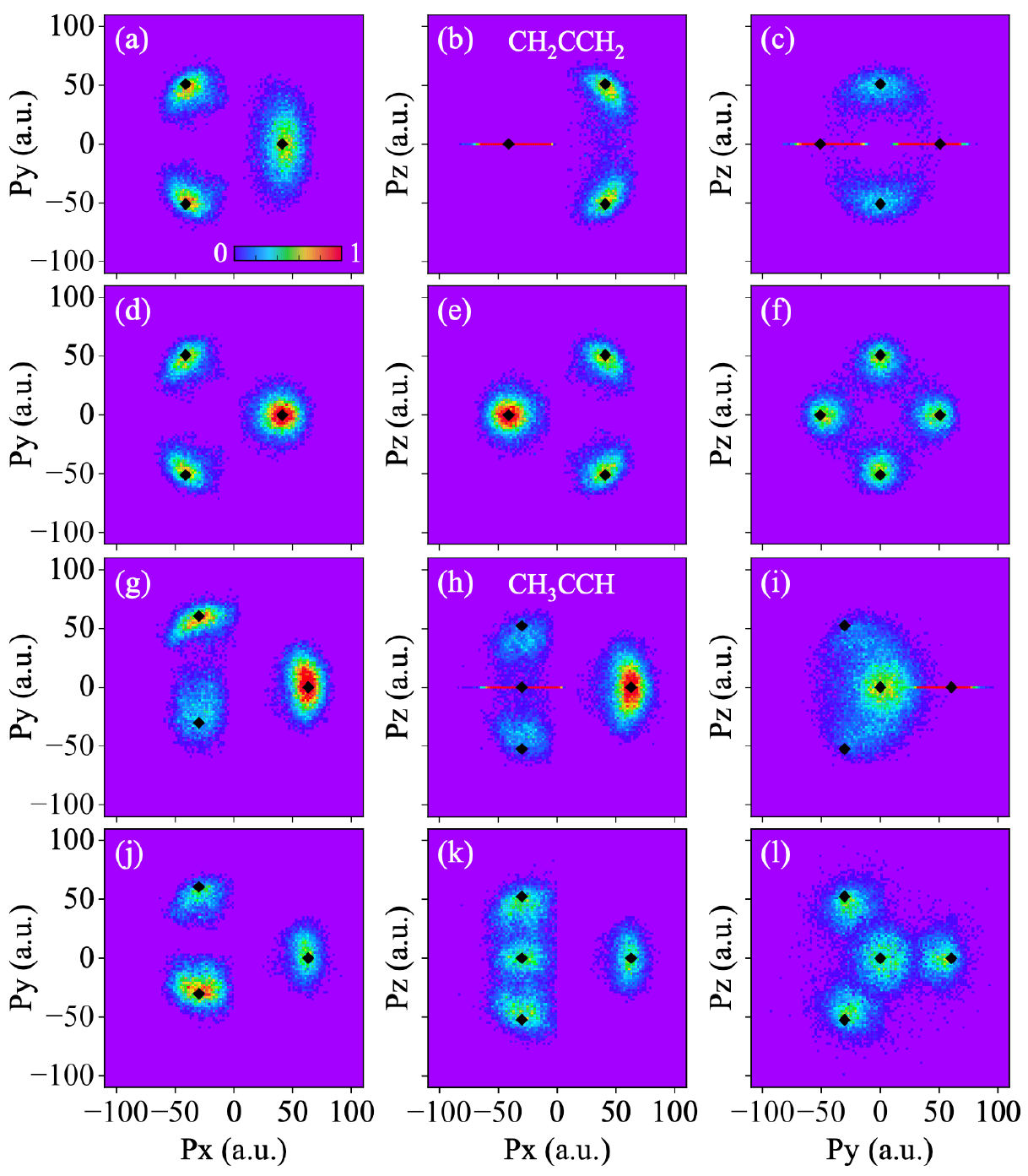}
		\caption{
Momentum imaging of hydrogen 3D configurations. (a)-(f) for CH$_2$CCH$_2$ and (g)-(l) for CH$_3$CCH. The left, middle, and right columns present projections onto xy, xz, and yz planes, respectively. The first and third rows display raw data in the Newton diagrams, while the second and fourth rows display data classified via k-means clustering. Black diamonds denote PC calculations. 
			\label{fig3}}
	\end{figure}
	%%%%%%%%%%%%%%%%%%%%%%%%%%%%%%%%%%%%%%%%%%%%%%%%%%%%%

The 3D momentum images of hydrogen are presented in Fig. 3, in which (a)-(c) display the Newton diagrams for allene in the xy, xz and yz planes, respectively. For the H$^+$ selected for definition of the xy plane, momentum component along z direction is inherently compressed to zero. In reality, this H should distribute similarly to the H at the other sites. We employ k-means clustering algorithm (see Supplemental Material Sec. V \cite{supp} and Refs. \cite{Li2022, Lloyd1982, MacQueen1967, Pedregosa2011} for details) to overcome this shortage in data presentation. The results are presented in Figs. 3(d)-(f), which show clear stereo images of all four H atoms for allene. Figures 3(g)-(i) present the momentum images for propyne. The two detected protons originating separately from the two terminal carbons dominate the 2H$^+$ + 3C$^+$ dissociation channel, accounting for 65\%, whereas both the two H$^+$ from CH$_3$ group accounts for the remaining 35\%. This explains why the H$^+$ from CH is much more intense than H$^+$ from CH$_3$, see Fig. 3(h) as an example. In order to maintain equivalent statistical counts, we randomly select 36\% of the events with H$^+$ from different terminals to image H$^+$ from CH group. For such events, only the proton from CH site is presented in Figs. 3(g)-(i), while the proton from the other site is only used for definition of the molecular axis. The events with both protons from the same terminal carbon are employed to image hydrogen of CH$_3$ group, and the k-means clustering procedure was applied only to such events (Supplemental Material Sec. V \cite{supp}). Figs. 3(j)-(l) present the data after k-means clustering procedure. These operations greatly improve the momentum imaging. For instance, features blurred in Fig. 3(i) are clearly resolved in Fig. 3(l). The maximums of the measured momentum distributions agree well with black diamonds from PC calculations for both isomers, revealing that a high-precision sub-angstrom resolution is achieved for H comparable to that for C.
	
CEI was envisioned to resolve molecular vibrational motions since its inception \cite{Vager1986, Vager1989}. However, only limited experiments have been reported, all restricted to small molecules \cite{Kitamura2001, Saito2004, Schmidt13}. Recently, Richard et al. investigated CEI of C$_5$H$_4$NI, in which the observed correlations between out-of-plane momentum components of different fragments were attributed to ground-state structural fluctuations  \cite{Richard2025}. The analysis in that work was restricted solely to out-of-plane, as in-plane components are strongly disturbed by complex dynamics of X-ray multi-photon ionization \textcolor{black}{and the followed Auger cascade}. Characterizing quantum fluctuations within the full 3D geometry remains challenging. The ultrahigh resolution ($\sim 0.1$ $\mathring{\mathrm{A}}$) and ultrafast ionization timescale ($< 1$ fs) \cite{Yuan2024} achieved with HCIs provide a robust basis for exploring ground-state fluctuations of the stereostructures. Here, we display such investigation for allene, while the analysis for propyne is given in Supplemental Material Sec. VI \cite{supp}.

	%%%%%%%%% fig 4 %%%%%%%%%%%%%%%%%%%%%%%%%%%%%%%%%%%%%
	\begin{figure}[tb]
		\centering
		\includegraphics[width=1 \linewidth]{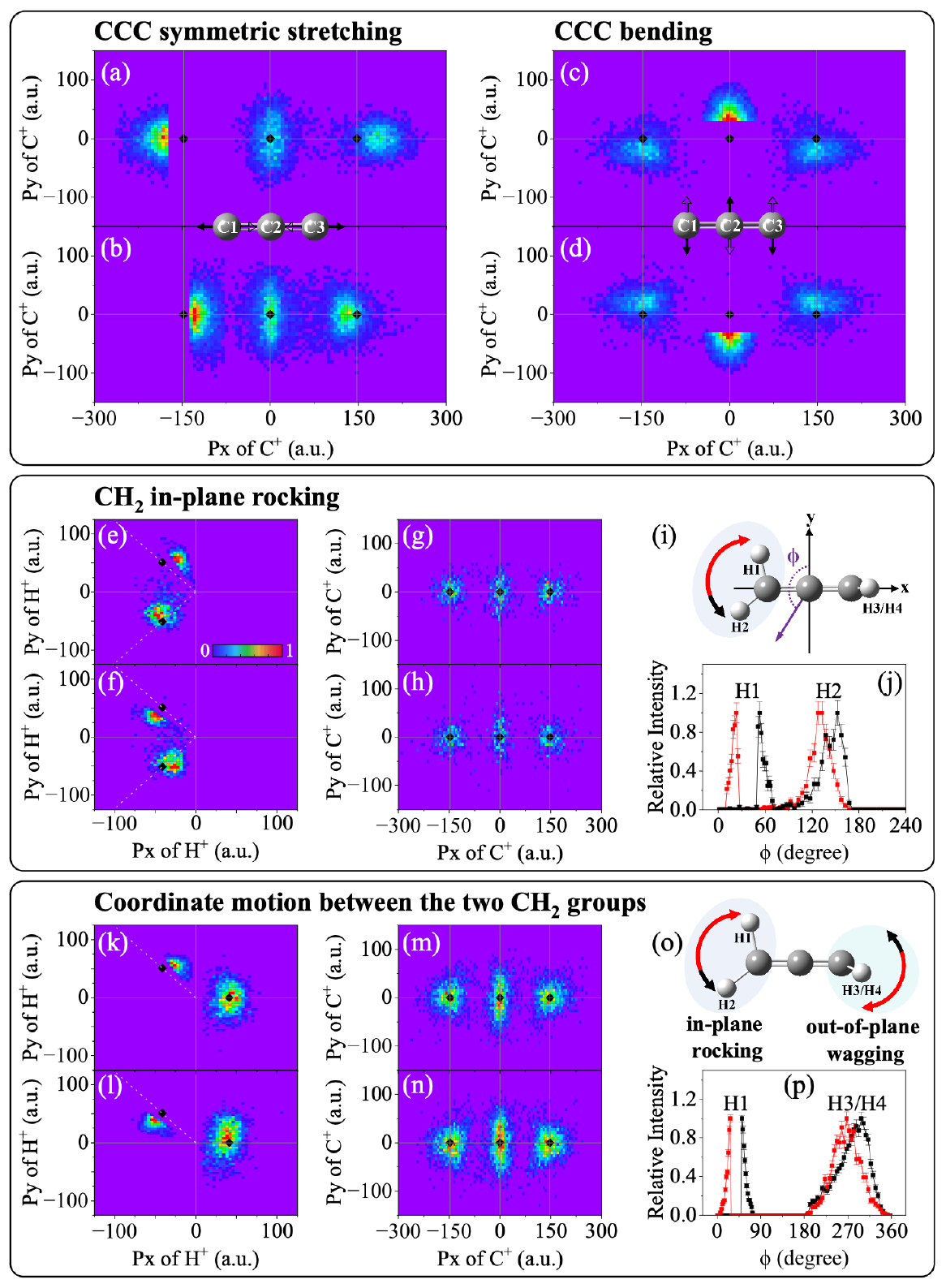}
		\caption{
3D structural fluctuations of ground-state CH$_2$CCH$_2$ probed via HCI-induced CEI. Top row: C$^+$ momentum correlations characterizing symmetric stretching (a, b) and bending (c, d) modes of the CCC skeleton. For symmetric stretching, filtering high-momentum signals for one terminal carbon yields enhanced momentum for carbon in the opposite end (a), and vice versa (b). For bending mode, central carbon emission along +y corresponds to enhanced -y momentum of two terminal carbons (c), and vice versa (d). Middle row: momentum correlation between two H$^+$ ions from one CH$_2$ group, revealing in-plane rocking motion. The two protons rotate synchronously in clockwise (e) and counterclockwise (f) directions. The azimuthal angle $\phi$ [defined in (i)] distributions in (j) further verify such synchronous angular variation. Bottom row: concerted motions between two spatially separated CH$_2$ groups. Counterclockwise in-plane rocking toward -y for the left CH$_2$ group is accompanied by counterclockwise out-of-plane wagging toward +y for the right CH$_2$ group (k, l), which is also confirmed by $\phi$ distributions in (p). Carbon momentum distributions (g, h, m, n) remain nearly unchanged under different H1 filters. Error bars represent standard deviations. Black diamonds denote results of PC calculations. \textcolor{black}{All experimental data shown are raw, without k-means clustering.} 
			\label{fig4}}
		
	\end{figure}
	%%%%%%%%%%%%%%%%%%%%%%%%%%%%%%%%%%%%%%%%%%%%%%%%%%%%%%	

The upper panels of Fig. 4 present momentum correlations among three C$^+$ ions. Figs. 4(a) and (b) depict the symmetric stretching mode of the CCC backbone. When a high-magnitude momentum filter is applied to the carbon at one end of the molecular skeleton, the carbon on the other end likewise shows higher momentum magnitudes [Fig. 4(a)], and the reverse also holds [Fig. 4(b)]. In addition, if the central carbon fragments along the +y direction, the two end carbons tend to carry larger momentum components along -y [Fig. 4(c)], and vice versa [Fig. 4(d)]. Such momentum correlations illustrate bending mode of the CCC skeleton.
	
We further investigate the collective motion of hydrogen atoms. The middle panels of Fig. 4 show momentum correlations between the two protons belonging to the same CH$_2$ group. These correlations uncover an in-plane rocking mode: the two protons move in concert, rotating either clockwise [Fig. 4(e)] or counterclockwise [Fig. 4(f)]. This is more clearly seen in the azimuthal angle ($\phi$) distributions, where $\phi$ is defined as the angle between momentum component of proton in the xy plane and the y-axis [Figs. 4(i)]. As shown in Fig. 4(j), the $\phi$ angle of H1 and H2 increase or decrease synchronously.
	
The bottom panels of Figs. 4(k)-(p) show momentum correlations between protons bound to different terminal carbons. The correlated motion of the two CH$_2$ moieties intervening by a central C is clearly uncovered. As elaborated above, the two hydrogen atoms within one CH$_2$ undergo concerted in-plane rocking. Accordingly, the momentum behavior of H1 in Figs. 4(k) and (l) serves as a marker for the in-plane rocking of the left-side CH$_2$ group. Both these correlation maps and the azimuthal angle distributions shown in Fig. 4(p) exhibit concerted dynamics for the two CH$_2$ groups at opposite ends. When the left CH$_2$ locating in the xy plane rocks counterclockwise toward -y from (k) to (l), the right CH$_2$ locating in the xz plane undergoes counterclockwise out-of-plane wagging toward +y, as sketched in Fig. 4(o).
\textcolor{black}{These plots distinctly reveal the collective motion that couples in-plane rocking of one CH$_2$ and out-of-plane wagging of the other CH$_2$.}
	
Meanwhile, the C$^+$ momentum distributions show no significant change for different proton momentum filters, as shown in Figs. 4(g), (h), (m) and (n). Similarly, the proton momentum distributions show no distinguishable variations with different C$^+$ momentum filters (see Supplemental Material Sec. VII \cite{supp}). The in-plane rocking and out-of-plane wagging are predominantly governed by the hydrogen atoms in CH$_2$, and have negligible influence on/from the CCC skeleton. In addition, the vibrational frequencies involving hydrogen differ substantially from those mainly involving carbon owing to their mass differences. Such vibration modes lie far from resonance, leading to dynamical decoupling of the respective vibrational degrees of freedom \cite{Fermi31}. Thus, the correlation between slow carbon skeletal motions and fast hydrogen motions is probably suppressed. This may explain why the momenta of H and C  exhibit no observable correlation.

In summary, we have demonstrated that a high resolution of 0.1 $\mathring{\mathrm{A}}$ can be achieved for C$_3$H$_4$ isomers by HCI-induced CEI. The high resolution achieved in this work not only enables to resolve the tiny length differences between CC single, double and triple bonds, but also provides a reliable tool to capture ground-state quantum fluctuations of the stereostructure in the full three dimensions. The quantum fluctuations corresponding to symmetric stretching and bending modes are clearly visualized by correlation analysis of the C$^+$ momenta. The momentum correlations between protons manifest the coupling between the in-plane rocking motion of one CH$_2$ group and the out-of-plane wagging of the other at the opposite end. Our work exhibits the ultrahigh resolution achieved by  HCI-based CEI, and opens new avenues for exploring 3D collective motions of multiple nuclei in stereostructures of molecules.
	
We thank the staff of 320-kV platform for multidiscipline research with HCIs for providing stable beams. The work was jointly supported by the National Key Research and Development Program of China under Grant No. 2022YFA1602500, the National Natural Science Foundation of China under Grant Nos. 12274416 and 12404308, and by Chinese Academy of Sciences.

 \bibliography{ref_CEI}

@article{Fermi31,
	Author = {Fermi, E. },
	Da = {1931/03/01},
	Doi = {10.1007/BF01341712},
	Id = {Fermi1931},
	Isbn = {0044-3328},
	Journal = {Zeitschrift f{\"u}r Physik},
	Number = {3},
	Pages = {250--259},
	Title = {{\"U}ber den Ramaneffekt des Kohlendioxyds},
	Ty = {JOUR},
	Url = {https://doi.org/10.1007/BF01341712},
	Volume = {71},
	Year = {1931}}

@misc{supp,
howpublished = {
See Supplemental Material for details of
experimental methods,
estimation of vibrational state distribution within the target,
definition of the molecular frame,
point-charge calculation,
k-means clustering procedure,
ground-state structural fluctuations of propyne, 
and proton momentum distributions under different carbon momentum filters.
}
}

@article{Schmidt13,
  title = {Spatial Imaging of the $\mathbf{H}_{2}{}^{+}$ Vibrational Wave Function at the Quantum Limit},
  author = {Schmidt, L. Ph. H. and Jahnke, T. and Czasch, A. and Sch\"offler, M. and Schmidt-B\"ocking, H. and D\"orner, R.},
  journal = {Phys. Rev. Lett.},
  volume = {108},
  issue = {7},
  pages = {073202},
  numpages = {4},
  year = {2012},
  month = {Feb},
  publisher = {American Physical Society},
  doi = {10.1103/PhysRevLett.108.073202},
  url = {https://link.aps.org/doi/10.1103/PhysRevLett.108.073202}
}

@article{Nellist2004,
	title   = {Direct Sub-Angstrom Imaging of a Crystal Lattice},
	author  = {P. D. Nellist and M. F. Chisholm and N. Dellby and O. L. Krivanek and M. F. Murfitt and Z. S. Szilagyi and A. R. Lupini and A. Borisevich and W. H. Sides and S. J. Pennycook},
	journal = {Science},
	volume  = {305},
	number  = {5691},
	pages   = {1741},
	year    = {2004},
	doi     = {10.1126/science.1100965},
	url = {https://www.science.org/doi/abs/10.1126/science.1100965}
}

@article{Jiang2018,
	title     = {Electron ptychography of 2D materials to deep sub-\aa ngstr\"om resolution},
	author    = {Jiang, Yi and Chen, Zhen and Han, Yimo and Deb, Pratiti and Gao, Hui and Xie, Saien and Purohit, Prafull and Tate, Mark W. and Park, Jiwoong and Gruner, Sol M. and Elser, Veit and Muller, David A.},
	journal   = {Nature},
	volume    = {559},
	number    = {7714},
	pages     = {343--349},
	year      = {2018},
	month     = {7},
	doi       = {10.1038/s41586-018-0298-5},
	url       = {https://doi.org/10.1038/s41586-018-0298-5}
}

@article{Blackburn2025,
	title     = {Sub-\aa ngstr\"om resolution ptychography in a scanning electron microscope at 20 keV},
	author    = {Blackburn, Arthur M. and Cordoba, Cristina and Fitzpatrick, Matthew R. and McLeod, Robert A.},
	journal   = {Nat. Commun.},
	volume    = {16},
	number    = {1},
	pages     = {8977},
	year      = {2025},
	month     = {10},
	day       = {14},
	doi       = {10.1038/s41467-025-64133-3},
	url       = {https://doi.org/10.1038/s41467-025-64133-3}
}

@article{Masar2024,
	title     = {Unveiling significant impact of subtle atomic displacement from equilibrium position in crystal lattice on electronic properties and photocatalytic activity of ZnO},
	author    = {Milan Masar and Hassan Ali and Ali Can Guler and Pavol Suly and Pavel Urbanek and Jan Antos and Barbora Hanulikova and Michal Machovsky and Ivo Kuritka},
	journal = {Mater. Today Chem.},
	volume    = {38},
	pages     = {102136},
	year      = {2024},
	doi       = {10.1016/j.mtchem.2024.102136},
	url       = {https://www.sciencedirect.com/science/article/pii/S2468519424002428}
}

@article{Richard2025,
	title     = {Imaging collective quantum fluctuations of the structure of a complex molecule},
	author    = {Beno\^it Richard  and Rebecca Boll  and Sourav Banerjee  and Julia M. Sch\"afer  and Zoltan Jurek  and Gregor Kastirke  and Kilian Fehre  and Markus S. Sch\"offler  and Nils Anders  and Thomas M. Baumann  and Sebastian Eckart  and Benjamin Erk  and Alberto De Fanis  and Reinhard D\"orner  and Sven Grundmann  and Patrik Grychtol  and Max Hofmann  and Markus Ilchen  and Max Kircher  and Katharina Kubicek  and Maksim Kunitski  and Xiang Li  and Tommaso Mazza  and Severin Meister  and Niklas Melzer  and Jacobo Montano  and Valerija Music  and Yevheniy Ovcharenko  and Christopher Passow  and Andreas Pier  and Nils Rennhack  and Jonas Rist  and Daniel E. Rivas  and Daniel Rolles  and Ilme Schlichting  and Lothar Ph. H. Schmidt  and Philipp Schmidt  and Daniel Trabert  and Florian Trinter  and Rene Wagner  and Peter Walter  and Pawel Ziolkowski  and Artem Rudenko  and Michael Meyer  and Robin Santra  and Ludger Inhester  and Till Jahnke },
	journal   = {Science},
	volume    = {389},
	number    = {6760},
	pages     = {650--654},
	year      = {2025},
	doi       = {10.1126/science.adu2637},
	url       = {https://www.science.org/doi/abs/10.1126/science.adu2637}
}

@article{Binnig1986,
	title     = {Atomic Force Microscope},
	author    = {Binnig, G. and Quate, C. F. and Gerber, Ch.},
	journal   = {Phys. Rev. Lett.},
	volume    = {56},
	pages     = {930--933},
	year      = {1986},
	doi       = {10.1103/PhysRevLett.56.930},
	url       = {https://link.aps.org/doi/10.1103/PhysRevLett.56.930}
}

@inbook{Williams1996,
	title     = {The Transmission Electron Microscope},
	author    = {Williams, David B. and Carter, C. Barry},
	booktitle = {Transmission Electron Microscopy: A Textbook for Materials Science},
	year      = {1996},
	publisher = {Springer US},
	address   = {Boston, MA},
	pages     = {3--17},
	doi       = {10.1007/978-1-4757-2519-3_1},
	url       = {https://doi.org/10.1007/978-1-4757-2519-3_1}
}

@article{Seiler1983,
	title     = {Secondary electron emission in the scanning electron microscope},
	author    = {Seiler, H},
	journal   = {J. Appl. Phys.},
	volume    = {54},
	number    = {11},
	pages     = {R1--R18},
	year      = {1983},
	doi       = {10.1063/1.332840},
	url       = {https://doi.org/10.1063/1.332840}
}

@article{Gross2009,
	title     = {The Chemical Structure of a Molecule Resolved by Atomic Force Microscopy},
	author    = {Leo Gross  and Fabian Mohn  and Nikolaj Moll  and Peter Liljeroth  and Gerhard Meyer},
	journal   = {Science},
	volume    = {325},
	number    = {5944},
	pages     = {1110--1114},
	year      = {2009},
	doi       = {10.1126/science.1176210},
	url       = {https://www.science.org/doi/abs/10.1126/science.1176210}
}

@article{Telychko2022,
	title     = {Sub-angstrom noninvasive imaging of atomic arrangement in 2D hybrid perovskites},
	author    = {Mykola Telychko  and Shayan Edalatmanesh  and Kai Leng  and Ibrahim Abdelwahab  and Na Guo  and Chun Zhang  and Jes\'us I. Mendieta-Moreno  and Matyas Nachtigall  and Jing Li  and Kian Ping Loh  and Pavel Jel\'inek  and Jiong Lu},
	journal   = {Sci. Adv.},
	volume    = {8},
	number    = {17},
	pages     = {eabj0395},
	year      = {2022},
	doi       = {10.1126/sciadv.abj0395},
	url       = {https://www.science.org/doi/abs/10.1126/sciadv.abj0395}
}

@article{Qiu2025,
	title     = {Deep learning for sub-\aa ngstr\"om-resolution imaging in uncorrected scanning transmission electron microscopy},
	author    = {Qiu, Zanlin and Meng, Yuan and Li, Junxian and Hong, Yanhui and Li, Ning and Han, Xiaocang and Liang, Yu and Cheng, Wing Ni and Ke, Guolin and Zhang, Linfeng and E, Weinan and Zhao, Xiaoxu and Zhang, Jin},
	journal   = {Natl. Sci. Rev.},
	volume    = {12},
	number    = {8},
	pages     = {nwaf235},
	year      = {2025},
	doi       = {10.1093/nsr/nwaf235},
	url       = {https://doi.org/10.1093/nsr/nwaf235}
}

@article{Hensley2012,
	title     = {Imaging of Isolated Molecules with Ultrafast Electron Pulses},
	author    = {Hensley, Christopher J. and Yang, Jie and Centurion, Martin},
	journal   = {Phys. Rev. Lett.},
	volume    = {109},
	pages     = {133202},
	year      = {2012},
	doi       = {10.1103/PhysRevLett.109.133202},
	url       = {https://link.aps.org/doi/10.1103/PhysRevLett.109.133202}
}

@article{Wolter2016,
	title     = {Ultrafast electron diffraction imaging of bond breaking in di-ionized acetylene},
	author    = {B. Wolter  and M. G. Pullen  and A.-T. Le  and M. Baudisch  and K. Doblhoff-Dier  and A. Senftleben  and M. Hemmer  and C. D. Schr\"oter  and J. Ullrich  and T. Pfeifer  and R. Moshammer  and S. Gr\"afe  and O. Vendrell  and C. D. Lin  and J. Biegert},
	journal   = {Science},
	volume    = {354},
	number    = {6310},
	pages     = {308--312},
	year      = {2016},
	doi       = {10.1126/science.aah3429},
	url       = {https://www.science.org/doi/abs/10.1126/science.aah3429}
}

@article{Yang2018,
	title     = {Imaging CF$_3$I conical intersection and photodissociation dynamics with ultrafast electron diffraction},
	author    = {Jie Yang  and Xiaolei Zhu  and Thomas J. A. Wolf  and Zheng Li  and J. Pedro F. Nunes  and Ryan Coffee  and James P. Cryan  and Markus G\"uhr  and Kareem Hegazy  and Tony F. Heinz  and Keith Jobe  and Renkai Li  and Xiaozhe Shen  and Theodore Veccione  and Stephen Weathersby  and Kyle J. Wilkin  and Charles Yoneda  and Qiang Zheng  and Todd J. Martinez  and Martin Centurion  and Xijie Wang},
	journal   = {Science},
	volume    = {361},
	number    = {6397},
	pages     = {64--67},
	year      = {2018},
	doi       = {10.1126/science.aat0049},
	url       = {https://www.science.org/doi/abs/10.1126/science.aat0049}
}

@article{Wolf2019,
	title     = {The photochemical ring-opening of 1,3-cyclohexadiene imaged by ultrafast electron diffraction},
	author    = {Wolf, T. J. A. and Sanchez, D. M. and Yang, J. and Parrish, R. M. and Nunes, J. P. F. and Centurion, M. and Coffee, R. and Cryan, J. P. and G\"uhr, M. and Hegazy, K. and Kirrander, A. and Li, R. K. and Ruddock, J. and Shen, X. and Vecchione, T. and Weathersby, S. P. and Weber, P. M. and Wilkin, K. and Yong, H. and Zheng, Q. and Wang, X. J. and Minitti, M. P. and Mart\'inez, T. J.},
	journal   = {Nat. Chem.},
	volume    = {11},
	pages     = {504--509},
	year      = {2019},
	doi       = {10.1038/s41557-019-0252-7},
	url       = {https://doi.org/10.1038/s41557-019-0252-7}
}

@article{Kupper2014,
	title     = {X-Ray Diffraction from Isolated and Strongly Aligned Gas-Phase Molecules with a Free-Electron Laser},
	author    = {K\"upper, Jochen and Stern, Stephan and Holmegaard, Lotte and Filsinger, Frank and Rouz\'ee, Arnaud and Rudenko, Artem and Johnsson, Per and Martin, Andrew V. and Adolph, Marcus and Aquila, Andrew and Bajt, Sa\v{s}a and Barty, Anton and Bostedt, Christoph and Bozek, John and Caleman, Carl and Coffee, Ryan and Coppola, Nicola and Delmas, Tjark and Epp, Sascha and Erk, Benjamin and Foucar, Lutz and Gorkhover, Tais and Gumprecht, Lars and Hartmann, Andreas and Hartmann, Robert and Hauser, G\"unter and Holl, Peter and H\"omke, Andre and Kimmel, Nils and Krasniqi, Faton and K\"uhnel, Kai-Uwe and Maurer, Jochen and Messerschmidt, Marc and Moshammer, Robert and Reich, Christian and Rudek, Benedikt and Santra, Robin and Schlichting, Ilme and Schmidt, Carlo and Schorb, Sebastian and Schulz, Joachim and Soltau, Heike and Spence, John C. H. and Starodub, Dmitri and Str\"uder, Lothar and Th\o{}gersen, Jan and Vrakking, Marc J. J. and Weidenspointner, Georg and White, Thomas A. and Wunderer, Cornelia and Meijer, Gerard and Ullrich, Joachim and Stapelfeldt, Henrik and Rolles, Daniel and Chapman, Henry N.},
	journal   = {Phys. Rev. Lett.},
	volume    = {112},
	pages     = {083002},
	year      = {2014},
	doi       = {10.1103/PhysRevLett.112.083002},
	url       = {https://link.aps.org/doi/10.1103/PhysRevLett.112.083002}
}

@article{Minitti2015,
	title     = {Imaging Molecular Motion: Femtosecond X-Ray Scattering of an Electrocyclic Chemical Reaction},
	author    = {Minitti, M. P. and Budarz, J. M. and Kirrander, A. and Robinson, J. S. and Ratner, D. and Lane, T. J. and Zhu, D. and Glownia, J. M. and Kozina, M. and Lemke, H. T. and Sikorski, M. and Feng, Y. and Nelson, S. and Saita, K. and Stankus, B. and Northey, T. and Hastings, J. B. and Weber, P. M.},
	journal   = {Phys. Rev. Lett.},
	volume    = {114},
	pages     = {255501},
	year      = {2015},
	doi       = {10.1103/PhysRevLett.114.255501},
	url       = {https://link.aps.org/doi/10.1103/PhysRevLett.114.255501}
}

@article{Stankus2019,
	title     = {Ultrafast X-ray scattering reveals vibrational coherence following Rydberg excitation},
	author    = {Stankus, Brian and Yong, Haiwang and Zotev, Nikola and Ruddock, Jennifer M. and Bellshaw, Darren and Lane, Thomas J. and Liang, Mengning and Boutet, S\'ebastien and Carbajo, Sergio and Robinson, Joseph S. and Du, Wenpeng and Goff, Nathan and Chang, Yu and Koglin, Jason E. and Minitti, Michael P. and Kirrander, Adam and Weber, Peter M.},
	journal   = {Nat. Chem.},
	volume    = {11},
	pages     = {716--721},
	year      = {2019},
	doi       = {10.1038/s41557-019-0291-0},
	url       = {https://doi.org/10.1038/s41557-019-0291-0}
}

@article{Ma2022,
	title     = {Ultrafast isolated molecule imaging without crystallization},
	author    = {Zhuoran Ma  and Xiao Zou  and Lingrong Zhao  and Fengfeng Qi  and Tao Jiang  and Pengfei Zhu  and Dao Xiang  and Jie Zhang},
	journal   = {Proc. Natl. Acad. Sci. U.S.A.},
	volume    = {119},
	number    = {15},
	pages     = {e2122793119},
	year      = {2022},
	doi       = {10.1073/pnas.2122793119},
	url       = {https://www.pnas.org/doi/abs/10.1073/pnas.2122793119}
}

@article{Odate2023,
	title     = {Brighter, faster, stronger: ultrafast scattering of free molecules},
	author    = {Asami Odate and Adam Kirrander and Peter M. Weber and Michael P. Minitti},
	journal   = {Adv. Phys.: X},
	volume    = {8},
	number    = {1},
	pages     = {2126796},
	year      = {2023},
	doi       = {10.1080/23746149.2022.2126796},
	url       = {https://doi.org/10.1080/23746149.2022.2126796}
}

@article{Vager1989,
	title     = {Coulomb Explosion Imaging of Small Molecules},
	author    = {Z. Vager  and R. Naaman  and E. P. Kanter},
	journal   = {Science},
	volume    = {244},
	number    = {4903},
	pages     = {426--431},
	year      = {1989},
	doi       = {10.1126/science.244.4903.426},
	url       = {https://www.science.org/doi/abs/10.1126/science.244.4903.426}
}

@article{Zhou2020,
	title     = {Coulomb explosion imaging for gas-phase molecular structure determination: An ab initio trajectory simulation study},
	author    = {Zhou, Weiwei and Ge, Lingfeng and Cooper, Graham A. and Crane, Stuart W. and Evans, Michael H. and Ashfold, Michael N. R. and Vallance, Claire},
	journal   = {J. Chem. Phys.},
	volume    = {153},
	number    = {18},
	pages     = {184201},
	year      = {2020},
	doi       = {10.1063/5.0024833},
	url       = {https://doi.org/10.1063/5.0024833}
}

@article{Yuan2024,
	title     = {Coulomb Explosion Imaging of Complex Molecules Using Highly Charged Ions},
	author    = {Yuan, Hang and Gao, Yue and Yang, Bo and Gu, Shaofei and Lin, Hong and Guo, Dalong and Liu, Junliang and Zhang, Shaofeng and Ma, Xinwen and Xu, Shenyue},
	journal   = {Phys. Rev. Lett.},
	volume    = {133},
	pages     = {193002},
	year      = {2024},
	doi       = {10.1103/PhysRevLett.133.193002},
	url       = {https://link.aps.org/doi/10.1103/PhysRevLett.133.193002}
}

@article{Boll2022,
	title     = {X-ray multiphoton-induced Coulomb explosion images complex single molecules},
	author    = {Boll, Rebecca and Sch\"afer, Julia M. and Richard, Beno\^it and Fehre, Kilian and Kastirke, Gregor and Jurek, Zoltan and Sch\"offler, Markus S. and Abdullah, Malik M. and Anders, Nils and Baumann, Thomas M. and Eckart, Sebastian and Erk, Benjamin and De Fanis, Alberto and D\"orner, Reinhard and Grundmann, Sven and Grychtol, Patrik and Hartung, Alexander and Hofmann, Max and Ilchen, Markus and Inhester, Ludger and Janke, Christian and Jin, Rui and Kircher, Max and Kubicek, Katharina and Kunitski, Maksim and Li, Xiang and Mazza, Tommaso and Meister, Severin and Melzer, Niklas and Montano, Jacobo and Music, Valerija and Nalin, Giammarco and Ovcharenko, Yevheniy and Passow, Christopher and Pier, Andreas and Rennhack, Nils and Rist, Jonas and Rivas, Daniel E. and Rolles, Daniel and Schlichting, Ilme and Schmidt, Lothar Ph. H. and Schmidt, Philipp and Siebert, Juliane and Strenger, Nico and Trabert, Daniel and Trinter, Florian and Vela-Perez, Isabel and Wagner, Rene and Walter, Peter and Weller, Miriam and Ziolkowski, Pawel and Son, Sang-Kil and Rudenko, Artem and Meyer, Michael and Santra, Robin and Jahnke, Till},
	journal   = {Nat. Phys.},
	volume    = {18},
	pages     = {423--428},
	year      = {2022},
	doi       = {10.1038/s41567-022-01507-0},
	url       = {https://doi.org/10.1038/s41567-022-01507-0}
}

@article{Lam2024,
	title     = {Differentiating Three-Dimensional Molecular Structures Using Laser-Induced Coulomb Explosion Imaging},
	author    = {Lam, Huynh Van Sa and Venkatachalam, Anbu Selvam and Bhattacharyya, Surjendu and Chen, Keyu and Borne, Kurtis and Wang, Enliang and Boll, Rebecca and Jahnke, Till and Kumarappan, Vinod and Rudenko, Artem and Rolles, Daniel},
	journal   = {Phys. Rev. Lett.},
	volume    = {132},
	pages     = {123201},
	year      = {2024},
	doi       = {10.1103/PhysRevLett.132.123201},
	url       = {https://link.aps.org/doi/10.1103/PhysRevLett.132.123201}
}

@article{Green2025,
	title     = {Visualizing the Three-Dimensional Arrangement of Hydrogen Atoms in Organic Molecules by Coulomb Explosion Imaging},
	author    = {Green, Alice E. and Chen, Keyu and Bhattacharyya, Surjendu and Allum, Felix and Usenko, Sergey and Ashfold, Michael N. R. and Baumann, Thomas M. and Borne, Kurtis D. and Brouard, Mark and Burt, Michael and Curchod, Basile F. E. and Erk, Benjamin and Forbes, Ruaridh J. G. and Ibele, Lea M. and Ingle, Rebecca A. and Van Sa Lam, Huynh and Li, Xiang and Lin, Kang and Mazza, Tommaso and McManus, Joseph W. and Meyer, Michael and Mullins, Terence and Figueira Nunes, Joao Pedro and Rivas, Daniel E. and Roerig, Aljoscha and Rouz\'ee, Arnaud and Schmidt, Philipp and Searles, John and Senfftleben, Bj\"orn and Stapelfeldt, Henrik and Tanyag, Rico Mayro P. and Trinter, Florian and Venkatachalam, Anbu Selvam and Wang, Enliang and Warne, Emily M. and Weber, Peter M. and Wolf, Thomas J. A. and Jahnke, Till and Rudenko, Artem and Boll, Rebecca and Rolles, Daniel},
	journal   = {J. Am. Chem. Soc.},
	volume    = {147},
	number    = {41},
	pages     = {37133--37143},
	year      = {2025},
	doi       = {10.1021/jacs.5c08730},
	url       = {https://doi.org/10.1021/jacs.5c08730}
}

@article{Cheng2026,
	title     = {Imaging Three-Dimensional Molecular Structure and Dynamics with Multiparticle Covariance and Cumulant Coulomb Explosion Analysis},
	author    = {Cheng, Chuan and Kumagai, Yoshiaki and Nagaya, Kiyonobu and Gejo, Tatsuo and Harries, James and Burt, Michael and Brouard, Mark and Duley, Avijit and Hockett, Paul and McManus, Joseph W. and Minns, Russell S. and Mondal, Subhendu and Owada, Shigeki and Razmus, Weronika and Rolles, Daniel and Sato, Takahiro and Thompson, Henry J. and Venkatachalam, Anbu S. and Warne, Emily M. and Walmsley, Tiffany and Yagi, Mana and Bucksbaum, Philip and Allum, Felix and Forbes, Ruaridh},
	journal   = {J. Phys. Chem. Lett.},
	volume    = {17},
	number    = {12},
	pages     = {3605-3616},
	year      = {2026},
    doi       = {10.1021/acs.jpclett.6c00162},
    url       = {https://doi.org/10.1021/acs.jpclett.6c00162},
}

@article{Richard2021,
	title     = {Statistical analysis of correlations in the x-ray induced Coulomb explosion of iodopyridine},
	author    = {Richard, Benoît and Sch\"afer, Julia M. and Jurek, Zoltan and Santra, Robin and Inhester, Ludger},
	journal   = {J. Phys. B: At. Mol. Opt. Phys.},
	volume    = {54},
	number    = {19},
	pages     = {194001},
	year      = {2021},
	doi       = {10.1088/1361-6455/ac2f4f},
	url       = {https://doi.org/10.1088/1361-6455/ac2f4f}
}

@article{Ashrafi2024,
	title     = {Reconstructing real-space geometries of polyatomic molecules undergoing strong field laser-induced Coulomb explosion},
	author    = {Ashrafi-Belgabad, Aydin and Karimi, Reza and Monfared, Mohammad and Tian, Kaili and Parvin, Parviz and Wales, Benji and Bisson, \'Eric and Beaulieu, Samuel and Gigu\'ere, Mathieu and Kieffer, Jean-Claude and Lassonde, Philippe and L\'egar\'e, Fran\c{c}ois and Ibrahim, Heide and Sanderson, Joseph H.},
	journal   = {Commun. Phys.},
	volume    = {7},
	pages     = {405},
	year      = {2024},
	doi       = {10.1038/s42005-024-01863-8},
	url       = {https://doi.org/10.1038/s42005-024-01863-8}
}

@article{Viveiros2026,
	title     = {Ab initio study of highly-charged-ion-induced Coulomb-explosion imaging},
	author    = {Viveiros, Misa and Taylor, Samuel S. and Covington, Cody and Varga, K\'alm\'an},
	journal   = {Phys. Rev. A},
	volume    = {113},
	pages     = {042805},
	year      = {2026},
	doi       = {10.1103/78dy-6wjq},
	url       = {https://link.aps.org/doi/10.1103/78dy-6wjq}
}

@article{Venkatachalam2025,
	title     = {Exploiting correlations in multi-coincidence Coulomb explosion patterns for differentiating molecular structures using machine learning},
	author    = {Venkatachalam, Anbu Selvam and Greenman, Loren and Stallbaumer, Joshua and Rudenko, Artem and Rolles, Daniel and Lam, Huynh Van Sa},
	journal   = {Nat. Commun.},
	volume    = {16},
	pages     = {11366},
	year      = {2025},
	doi       = {10.1038/s41467-025-66369-5},
	url       = {https://doi.org/10.1038/s41467-025-66369-5}
}

@article{Guo2025,
	title         = {Decoding Molecular Geometries in Coulomb Explosion Imaging via Physics-Informed Deep Neural Network},
	author        = {Xingyu Guo and Enliang Wang and Wenguang Wu and Zhaopeng Xing and Tuo Liu and Chunkai Xu and Xu Shan and Artem Rudenko and Daniel Rolles and Jing Chen and Xiangjun Chen},
	eprint        = {2512.16559},
	archivePrefix = {arXiv},
	url           = {https://arxiv.org/abs/2512.16559}
}

@article{Li2026,
	title     = {Generative modeling enables molecular structure retrieval from Coulomb explosion imaging},
	author    = {Li, Xiang and Jahnke, Till and Boll, Rebecca and Han, Jiaqi and Xu, Minkai and Meyer, Michael and Piancastelli, Maria Novella and Rolles, Daniel and Rudenko, Artem and Trinter, Florian and Wolf, Thomas J. A. and Thayer, Jana B. and Cryan, James P. and Ermon, Stefano and Ho, Phay J.},
	journal   = {Nat. Commun.},
	volume    = {17},
	pages     = {3430},
	year      = {2026},
	doi       = {10.1038/s41467-026-70160-5},
	url       = {https://doi.org/10.1038/s41467-026-70160-5}
}

@article{Andre2025,
	title     = {Protein Structure Classification Based on X-Ray-Laser-Induced Coulomb Explosion},
	author    = {Andr\'e, Tomas and Dawod, Ibrahim and Cardoch, Sebastian and De Santis, Emiliano and T\^imneanu Nicu\c{s}or and Caleman, Carl},
	journal   = {Phys. Rev. Lett.},
	volume    = {134},
	pages     = {128403},
	year      = {2025},
	doi       = {10.1103/PhysRevLett.134.128403},
	url       = {https://link.aps.org/doi/10.1103/PhysRevLett.134.128403}
}

@article{Ullrich2003,
	title     = {Recoil-ion and electron momentum spectroscopy: reaction-microscopes},
	author    = {J Ullrich and R Moshammer and A Dorn and R D\"orner and L Ph H Schmidt and H Schmidt-B\"ocking},
	journal   = {Rep. Prog. Phys.},
	volume    = {66},
	number    = {9},
	pages     = {1463},
	year      = {2003},
	doi       = {10.1088/0034-4885/66/9/203},
	url       = {https://doi.org/10.1088/0034-4885/66/9/203}
}

@article{Dorner2000,
	title     = {Cold Target Recoil Ion Momentum Spectroscopy: a ‘momentum microscope’ to view atomic collision dynamics},
	author    = {R. D\"orner and V. Mergel and O. Jagutzki and L. Spielberger and J. Ullrich and R. Moshammer and H. Schmidt-B\"ocking},
	journal   = {Phys. Rep.},
	volume    = {330},
	number    = {2},
	pages     = {95--192},
	year      = {2000},
	doi       = {10.1016/S0370-1573(99)00109-X},
	url       = {https://doi.org/10.1016/S0370-1573(99)00109-X}
}

\end{document}